\documentclass[10pt,aps,amsmath,amssymb,showpacs,floatfix,twocolumn]{revtex4}
\usepackage[draft]{hyperref}
\usepackage{amsmath}
\usepackage{graphicx}
\begin{document}
\title{Cavity-enhanced absorption for optical refrigeration}
\author{Denis V. Seletskiy$^{*}$, Michael P. Hasselbeck, and Mansoor Sheik-Bahae}
\affiliation{Department of Physics and Astronomy, University of New Mexico, \\ 800 Yale Blvd NE, NM, Albuquerque,
87131, USA \\ $^{*}$Corresponding author: denisel@unm.edu}
\date{\today}
\begin{abstract}
A 20-fold increase over the single path optical absorption is demonstrated with a low loss medium placed in a resonant
cavity.  This has been applied to laser cooling
of Yb:ZBLAN glass resulting in 90\% absorption of the incident pump light.  A coupled-cavity scheme to
achieve active optical impedance matching is analyzed.
\end{abstract}

\maketitle

\noindent  The benefits of absorption enhancement in an optical cavity were appreciated shortly after the invention of
the laser \cite{Kastler1962AO}.  The effective interaction length increases as a consequence of beam trapping inside a stable resonator.  Techniques such as cavity ring-down spectroscopy exploit this effect to achieve $10^{-10}$
absorption sensitivity in gaseous media \cite{OKeefe1988RSI,Ye2000PRA}. Broadband spectroscopy can be performed with
the cavity ring-down concept by means of an optical frequency comb \cite{Thorpe2006Sci}.  The high sensitivity achieved with gases is possible because of the low intrinsic absorption and minimal parasitic losses.

Use of an optical cavity in condensed
matter spectroscopy is usually not necessary because of the larger absorption.  There are applications, however, that can benefit from placing a weakly absorbing solid-state medium in an optical resonator.
Resonantly enhanced absorption has been shown to increase the electronic bandwidth of infrared detectors and provides
spectral tuning of the response~\cite{Unlu1}.  Here we use an optical cavity to increase pump light absorption in
experiments to optically refrigerate glass.

Laser cooling in solids occurs by anti-Stokes luminescence,
i.e.~conversion of coherent pump photons into isotropic, higher
energy luminescence photons\cite{Pringsheim1929,Landau1946}.
The energy gained in the luminescence process is supplied by
phonon absorption. If excited state relaxation is predominantly
radiative, net cooling of the high purity medium occurs. In
1995, Epstein \emph{et al} demonstrated net laser cooling in
solids for the first time by using high purity fluoro-zirconate
glass (ZBLAN) that had minimal non-radiative losses
\cite{Epstein1995Nat}. Since then, a variety of solid hosts,
both amorphous and crystalline, doped with rare-earth ions of
ytterbium (Yb), thulium, and erbium have been cooled by laser
light \cite{MSB2007Nat,MSB2009Book}. The cooling power (i.e.~rate of heat
lift) of an optical refrigerator is $P_{\text{cool}} =
P_{\text{abs}}(\eta \nu_{\text{F}}/\nu - 1)$, where
$P_{\text{abs}}$ is the absorbed optical power, $\nu$ is
the pump frequency, $\nu_{\text{F}}$ is the mean
fluorescence frequency, and $\eta$ is the external quantum
efficiency that accounts for photon emission and escape
\cite{Mungan1997PRL}. To attain net cooling
(i.e.~$P_{\text{cool}} > 0$), pumping must take place at
frequencies smaller than the mean luminescence frequency. This
is where absorption is inherently weak, however, which makes
$P_{\text{abs}}$ very low.

Pump absorption can be improved by increasing the interaction
length with a multi-pass scheme. A non-resonant trapping
geometry places the sample between two dielectric mirrors with
pump light admitted through a small entrance hole in the input
mirror. This approach led to cooling by 90 degrees starting
from room temperature in Yb:ZBLAN glass \cite{Thiede2005APL}.
Much lower temperatures have been recently attained with a
Yb-doped yttrium-lithium fluoride crystal host using both
resonant and non-resonant absorption enhancement
\cite{dvs2008SPIE,dvs2009CLEO}.

\begin{figure}[htb]
\includegraphics[viewport=3 36 582 312,clip,width=8.3cm]{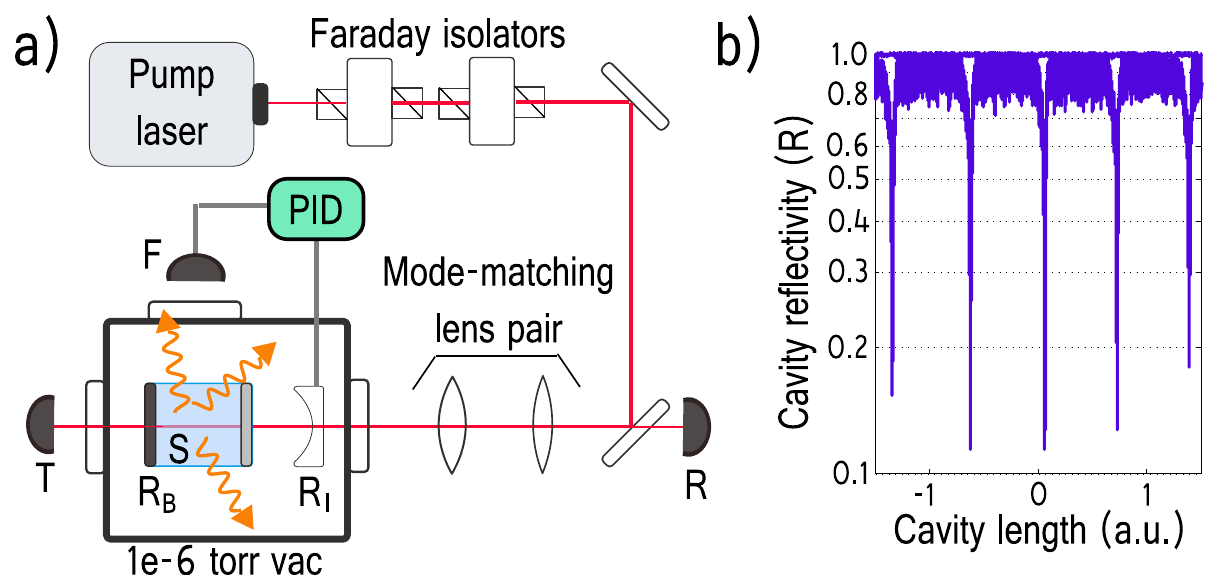}  
\caption{\label{fig:setup} a) Setup for RCE laser cooling.  Abbreviations are explained in the text; b)
Resonant 89\% reflectivity of the cavity plotted on a semi-log scale.}
\end{figure}

We use an optical cavity to dramatically enhance pump light absorption in a Yb:ZBLAN glass sample.
Resonant cavity enhancement (RCE) offers the benefits of increased interaction length without complications associated with the entrance
hole in a multi-pass setup, i.e.~pump light scatter and heating, as well as pump light leakage from the trap.
The sample is positioned between two mirrors to form a stable cavity as shown in Fig.~\ref{fig:setup}a.  The maximum absorption of the resonant cavity occurs when the reflectivity of the input mirror (\emph{I}) satisfies
$R_{\text{I}} = R_{\text{B}}\exp(-2\alpha L)$, where $R_{\text{B}}$ is the back mirror reflectivity, and $\alpha$ is the absorption coefficient of the glass sample of length $L$ ~\cite{Unlu1}.  Parasitic (background) loss is ignored.
The equation expresses optical impedance matching for a lossy cavity \cite{Siegman}.

In a Gires-Tournois cavity geometry, the back mirror reflectivity $R_{\text{B}}$ is taken as unity and the pump absorption on
resonance is given by

\begin{equation}
A = 1 - \left(\frac{\sqrt{R_{\text{I}}} - e^{-\alpha L}}{1 - e^{-\alpha L}\sqrt{R_{\text{I}}}}\right)^2.
\label{eqn:Abs}
\end{equation}

\noindent When the optical impedance matching condition is satisfied in the Gires-Tournois limit, absorption goes to unity.

The sample used in this work is much longer than a pump wavelength,
i.e.~$ 2 n L / \lambda \gg 1$ so positioning in the cavity
is not critical. In a thin detector application, where the
active element is of the order of a wavelength thick, it must
be located at an anti-node of the cavity standing wave
\cite{Jevrase}. We also note that the cavity pumping scheme can
be implemented by placing an optically cooled sample directly
inside a laser resonator~\cite{Heeg2004PRA}. This approach is
complicated by the fact that the temperature-dependent
round-trip loss of the sample will affect the lasing threshold.

Referring to Figure \ref{fig:setup}a, the pump field is derived
from a commercial continuous-wave Yb:YAG disk laser (ELS
Versadisk) nominally oscillating at 1030 nm.
After passing through a pair of Faraday isolators (60 dB
combined rejection) and spatial mode-matching lens pair, the
pump beam enters an optical cavity placed inside a vacuum
chamber. The cooling sample (\emph{S}) is fluoro-zirconate
glass ZBLAN (ZrF$_{4}$-BaF$_{2}$-LaF$_{3}$-AlF$_{3}$-NaF) doped
with 2\% w.t.~Yb$^{3+ }$. It is cylindrical in shape, length:~1
cm length, and diameter:~1 cm, which is similar to the sample
used in Ref.~\cite{Thiede2005APL}. A high-quality dielectric
mirror is deposited on the back facet; the front facet is
anti-reflection coated at the pump wavelength ($R_{\text{AR}} <
0.2\%$). The sample is supported by quartz optical fibers to
minimize conductive heat load from the environment. The
radiative load can be reduced by nearly 6 times by placing the
sample inside a shell coated with low thermal emissivity film
on its interior surfaces \cite{Thiede2005APL}. In this work, a
partial shell is used to allow thermal camera access for
non-contact temperature measurements described below. We
estimate the partial shell provides about half the radiation
shielding of a full shell. An input coupling mirror (\emph{I})
of reflectivity $R_{\text{I}}= 94\%$ is housed in a 3-axis
piezo-actuated mount, allowing for cavity length scan and
stabilization. This reflectivity is chosen to satisfy optical
impedance matching with the sample starting at room
temperature. The transmission of the back mirror is
$T_{\text{B}} < 0.1\%$, consistent with the Gires-Tournois
geometry. Losses at both $T_{\text{B}}$ and $R_{\text{AR}}$ are
negligible compared to the sample single-pass absorbance
$\left(4.5 \pm 0.5\%\right)$, ensuring that absorption
enhancement is largely in the sample. For the measured values
of $R_{\text{I}}$ and $\alpha L$, the calculated on-resonance
absorption (Eq.~\ref{eqn:Abs}) lies in the range $96.5\pm 2\%$.

A reflectivity signal (\emph{R}) is used to estimate the resonant absorption of the Gires-Tournois
cavity. Normalization is obtained from a high-reflectivity dielectric mirror placed immediately before the input coupler ($R_{I}$).  Since transmission is negligible in the Gires-Tournois geometry, we deduce absorption $A$ from
energy conservation: $A = 1-R$ while ignoring interface and scattering losses.  The experimental on-resonance
absorption of $89 \pm 3\%$ is shown in Fig.~\ref{fig:setup}b. This result is in good agreement with
the theoretical prediction, which is reduced from the nominal value of $96.5\pm 2\%$ by imperfect mode-matching as
evidenced by fringe asymmetry in the reflectivity signal.

The cavity is kept on resonance with an active stabilization scheme.  Laser-induced isotropic luminescence is fiber-coupled into a photo detector (\emph{F}) and used as an error-generating signal for the feedback loop.  The cavity length is modulated by a small-amplitude voltage from the internal reference of a lock-in amplifier. By mixing luminescence and dither voltage signals in a lock-in filter it is possible to differentiate the leading and trailing edges of a cavity fringe.  This allows generation of an unambiguous directional feedback signal via a proportional-integral-derivative (\emph{PID}) circuit (SRS, Model SIM960).

\begin{figure}[htb]
\centerline{\includegraphics[bb=68 140 370 310,clip,width=8.3cm]{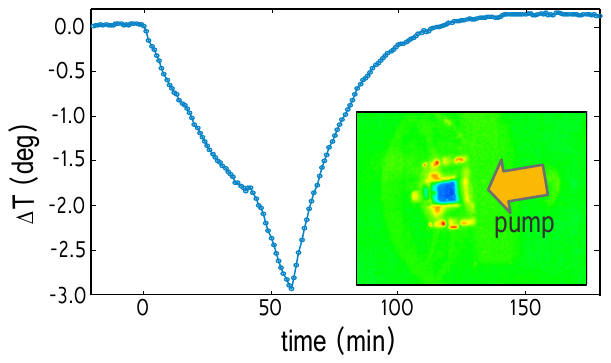}}  
\caption{\label{fig:Temp} Temperature evolution of RCE laser cooling.  Pump starts at
t = 0; blocked at t = 60 min;  Inset: Top view thermal image of cavity on
resonance.  Sample (blue) cools with respect to the hotter thermal radiation shell (red/yellow).}
\end{figure}
\bigskip

Laser cooling experiments are performed by irradiating the sample with 2 Watts of pump laser light. The black-body
radiation emitted by the sample is imaged onto a micro-bolometer thermal camera, thus providing a non-contact scene
temperature measurement. Averaged pixel counts from a time series of thermal images are converted to temperature by
means of a calibration coefficient obtained in a separate experiment.  The time evolution of temperature is presented
in Figure~\ref{fig:Temp}.  A thermal image is shown in the inset corresponding to resonant cooling by 3 degrees below
the temperature of the adjacent radiation shell.  The cooling process is terminated by blocking the laser 60 minutes
into the experiment to avoid saturation of the thermal camera image.  Exponential fitting of the cooling data
extrapolates to a steady state temperature decrease of 5 degrees below the shell temperature.

When account is made for non-ideal experimental conditions, our results compare favorably with the nearly
45 degrees of cooling reported at the same pump power and in a non-resonant arrangement \cite{Thiede2005APL}.  If we adjust
absorption efficiency (90\%), current cavity stabilization efficiency (60\%), and radiation shielding (30\%) to their
ideal values, a temperature drop of 30~K from ambient is predicted. Stabilization efficiency is limited by
longitudinal mode instability of the pump laser. Sample-to-sample impurity variations can account
for the remaining performance discrepancy~\cite{Hehlen2007PRB}.

An important advantage of RCE compared to non-resonant trapping schemes is its scalability.  Cavity
mode-volume and hence sample size can be matched in an RCE absorption arrangement.  The dominant parasitic heat load
is radiative, which scales linearly with sample surface area. Reducing the mode-volume along with sample dimensions
will improve laser cooling performance.  An added benefit will be miniaturization of the cryocooler.

The above proof-of-principle experiment demonstrates the
feasibility of RCE in laser cooling of solids. For large
temperature excursions (i.e.~to reach cryogenic temperatures),
dynamic optical impedance matching becomes an important
consideration. Due to the temperature-dependent absorption
coefficient, total absorption $A$ in a fixed cavity can be
maximum only at a particular temperature. Optimal coupling
requires that the input mirror reflectivity $R_{\text{I}}$
change to accommodate the change in the intra-cavity loss. A
work-around solution is to under-couple the cavity at room
temperature so as to satisfy impedance matching at a lower,
steady-state temperature $T^{'}$, where ~$\alpha(T^{'})$ is
known. A more general solution is to implement a continuously
tunable reflectivity. This concept was recently demonstrated in
fiber-based cavity, where input reflectivity was tuned by an
adjustable evanescent coupling loss in the input
fiber~\cite{Chow2008OE}. Here, we propose a free-space optics
solution in the form of a coupled cavity geometry to allow for
active impedance matching.

Without loss of generality, consider a symmetric cavity of reflectivity $R_{1}$ followed by an absorbing sample and a
third mirror $R_{2}$.  The half-round trip phases of the first (coupling) and second (absorbing) sub-cavities are $\phi_{1,2}$. Within the adiabatic approximation~\cite{Lang1986PRA}, the optical impedance matching condition can be generalized to:

\begin{equation}
R_{11}\left(\phi_{1}\right) = \frac{F \sin^{2}\left(\phi_{1}\right)}{1 + F \sin^{2}\left(\phi_{1}\right)} = R_{2}e^{-2\alpha L},
\label{eqn:OIMnew}
\end{equation}

\noindent where $F = 4R_{1}/(1-R_{1})^{2}$. Coupling between the two cavities causes Eq.~(\ref{eqn:OIMnew}) to be
satisfied only for a particular value of $\phi_2$.  Resonant reflectivity of the first (coupling) cavity ($R_{11}$)
acts as a tunable input coupler to maximize the absorption.  The maximum reflectivity
given by Eq.~(\ref{eqn:OIMnew}) sets the minimum absorbance that allows impedance matching:

\begin{equation}
(\alpha L)_{\text{min}} = \ln \left( \frac{1+R_{1}}{2\sqrt{R_{1}}}\right) 
\label{eqn:alphaMIN}
\end{equation}
\noindent for $R_{2} = 1$. The case of interest here involves small values of $\alpha L$, so the minimum reflectivity $R_{1}$ that satisfies the optical impedance matching condition is $R_{1} \approx 1 - \sqrt{8 \alpha L}$. This means that a range of intra-cavity loss values can be actively impedance matched by the input coupling cavity. To satisfy optical impedance matching for $\alpha L \sim 10^{-9}$, a reflectivity of $R_{1} \geq 0.9999$ is required. When the resonant absorption is extremely small, however, losses at the cavity mirrors can no longer be ignored in the analysis.

The proposed technique of optical impedance matching via a coupled cavity geometry is general.  We envision several
potential applications outside of laser cooling in solids.  Background-free, narrow-band, ultra-sensitive spectroscopy
is possible, where an acoustic signal proportional to weak absorbance can be monitored as a function of sub-cavity
phases.

In summary, we used a resonant cavity to obtain nearly 90\% absorption in a laser cooling sample, corresponding to
20-fold enhancement over single-pass absorbance.  A 5 degree temperature drop from the ambient was shown to be
comparable with the performance of state-of-the-art glass-host coolers when accounting for experimental inefficiencies.
We propose and outline a coupled cavity scheme to achieve active optical impedance-matching in high-power laser cooling experiments.



\end{document}